\documentstyle[12pt,fleqn]{article}
\def\singlespace {\smallskipamount=3.75pt plus1pt minus1pt
                  \medskipamount=7.5pt plus2pt minus2pt
                  \bigskipamount=15pt plus4pt minus4pt
                  \normalbaselineskip=15pt plus0pt minus0pt
                  \normallineskip=1pt
                  \normallineskiplimit=0pt
                  \jot=3.75pt
                  {\def\smallskip {\vskip\smallskipamount}}
                  {\def\medskip   {\vskip\medskipamount}}
                  {\def\bigskip   {\vskip\bigskipamount}}
                  {\setbox\strutbox=\hbox{\vrule 
                    height10.5pt depth4.5pt width 0pt}}
                  \parskip 7.5pt
                  \normalbaselines}
\def\middlespace {\smallskipamount=5.625pt plus1.5pt minus1.5pt
                  \medskipamount=11.25pt plus3pt minus3pt
                  \bigskipamount=22.5pt plus6pt minus6pt
                  \normalbaselineskip=22.5pt plus0pt minus0pt
                  \normallineskip=1pt
                  \normallineskiplimit=0pt
                  \jot=5.625pt
                  {\def\smallskip {\vskip\smallskipamount}}
                  {\def\medskip   {\vskip\medskipamount}}
                  {\def\bigskip   {\vskip\bigskipamount}}
                  {\setbox\strutbox=\hbox{\vrule 
                    height15.75pt depth6.75pt width 0pt}}
                  \parskip 11.25pt
                  \normalbaselines}
\def\doublespace {\smallskipamount=7.5pt plus2pt minus2pt
                  \medskipamount=15pt plus4pt minus4pt
                  \bigskipamount=30pt plus8pt minus8pt
                  \normalbaselineskip=30pt plus0pt minus0pt
                  \normallineskip=2pt
                  \normallineskiplimit=0pt
                  \jot=7.5pt
                  {\def\smallskip {\vskip\smallskipamount}}
                  {\def\medskip   {\vskip\medskipamount}}
                  {\def\bigskip   {\vskip\bigskipamount}}
                  {\setbox\strutbox=\hbox{\vrule 
                    height21.0pt depth9.0pt width 0pt}}
                  \parskip 15.0pt
                  \normalbaselines}

\include{dspace12}
\def\be{\begin{equation}}
\def\ee{\end{equation}}
\def\bea{\begin{eqnarray}}
\def\eea{\end{eqnarray}}
\def\nn{\nonumber}
\def\th{\theta}

\def\lt{\left}
\def\rt{\right}
\def\sect #1{\setcounter{equation}{0}}

\begin{document}
\singlespace
\vspace{0.5in}

\begin{center}
{\LARGE { Energy associated with  charged dilaton black holes}}
\end{center}
\vspace{0.4in}
\begin{center}
{\large{A. Chamorro\footnote[1]{E-mail\ :\ wtpchbea@lg.ehu.es} 
  and K. S. Virbhadra\footnote[2]{Present address : Theoretical Astrophysics
  Group, Tata Institute of Fundamental Research, Homi Bhabha
Road, Colaba, Bombay 400005, India; E-mail : shwetketu@tifrvax.tifr.res.in}\\
Departamento de F\'{\i}sica Te\'{o}rica\\
Universidad del Pa\'{\i}s Vasco\\
Apartado 644, 48080 Bilbao, Spain\\
}}
\end{center}
\vspace{0.8in}
\begin{abstract}
It is known that certain properties of  charged dilaton black holes
depend on a free parameter $\beta$ which controls the strength of the 
coupling of the dilaton to the Maxwell field. We obtain the energy
associated with static spherically symmetric charged dilaton black holes
for arbitrary value of the coupling parameter and find that the energy
distribution depends on the value of $\beta$. With  increasing radial
 distance, the energy in a sphere increases for
 $\beta = 0$ as well as for $\beta < 1$, decreases
for $\beta > 1$, and remains constant for $\beta = 1$.
However, the total  energy turns out
to be the same  for all values of $\beta$.
\end{abstract}
\vspace{1.0in}
\begin{center}
{\it  To appear in Int. J. Mod. Phys. D }
\end{center}
\newpage
Since the last few years there has been a growing interest in obtaining dilaton
black hole solutions and studying their properties which has led to several new 
insights about black holes ( \cite{GM88}-\cite{Dow94}). Garfinkle, Horowitz,
 and Strominger(GHS)\cite{GHS91}  obtained a nice form of  static
spherically symmetric charged dilaton black hole solutions which exhibit 
several different properties compared to the Reissner-Nordstr\"{o}m (RN) black 
holes. The dilaton has an important role in these solutions and many of their
interesting features have their origin in the  coupling between the Maxwell
and the dilaton fields. GHS considered the action (\cite{GHS91}-\cite{HH92})
\be
S\ =\ \int d^4 x \ \sqrt{-g} \ \lt[\ - R \ +\ 2 {(\nabla\Phi)}^2\ +
\ e^{-2\beta \Phi} \ F^2 \rt] ,
\ee
where $R$ is the Ricci scalar, $\Phi$ is the dilaton field, $F_{ab}$ is the
electromagnetic field tensor, and $\beta$ is a dimensionless free parameter 
which regulates the coupling between the dilaton and the Maxwell fields.
It is clear that a change in the sign of the free parameter $\beta$ is the same
as a change in the sign of the dilaton field $\Phi$;  therefore 
it is adequate to discuss only  nonnegative values of $\beta$.
$(1)$ gives the action
 for the Einstein-Maxwell-scalar theory for $\beta = 0$ and the action for 
the Kaluza-Klein theory for $\beta = \sqrt{3}$. Moreover, $\beta = 1$ in
$(1)$ gives the action which is a part of the low-energy action of  string
theory.
Varying $(1)$ gives the equations of motion:\cite{HH92}
\bea
& &\nabla_i \lt(e^{-2\beta\Phi} F^{ik}\rt) = 0 , \nn\\
& &{\nabla}^2\Phi + \frac{\beta}{2} e^{-2\beta \Phi}\ F^2 = 0 ,  \nn\\
& &R_{ik} = 2 \nabla_i\Phi \nabla_k\Phi + 2 e^{-2\beta\Phi} F_{ia} F_k^{\ a}
        -\frac{1}{2} g_{ik} e^{-2\beta\Phi}\ F^2 \ .
\eea
GHS obtained  static spherically symmetric charged dilaton black hole solutions
of  these equations  given by the line element,
\be
ds^2 = B dt^2 - A dr^2 - D r^2 (d\th^2 + \sin^2\th d\varphi^2),
\ee
the dilaton field $\Phi$ 
\be
e^{2\Phi} = \lt[1-\frac{r_{-}}{r}\rt]^{(1-\sigma)/\beta},
\ee
and the electromagnetic field tensor component
\be
F_{tr} = \frac{Q}{r^2},
\ee
where
\be
B = A^{-1} = \lt(1-\frac{r_{+}}{r}\rt) \lt(1-\frac{r_{-}}{r}\rt)^{\sigma},
\ee
\be
D = \lt(1-\frac{r_{-}}{r}\rt)^{1-\sigma},
\ee
and
\be
\sigma\ = \ \frac{1-{\beta}^2}{1+{\beta}^2} \ .
\ee
$r_{+}$ and $r_{-}$ are  related to the mass and charge
parameters, M and Q, 
through 
\bea
& &2M = r_{+}\ +\ \sigma\ r_{-} ,  \nn\\
& &Q^2 \lt(1+\beta^2\rt) = r_{+}\ r_{-} \ .
\eea
For $\beta = 0$ the solution yields the well known RN
solution of the Einstein-Maxwell equations. 
The surface $r = r_{+}$ is an event horizon for all values of $\beta$.

There are several properties of  charged dilaton black holes which depend
crucially  on the free parameter $\beta$. The maximum charge which can
be carried by a charged dilaton black hole (for a given mass) depends on the
value of $\beta$ (see \cite{HW92}). However, for a given mass there exits an extremal
limit for all values of $\beta$.
When $\beta \neq 0$, the surface
 $r = r_{-}$ is a curvature singularity  and when $\beta
=0$ (the case of the RN solution) the surface $r = r_{-}$
is a nonsingular inner horizon\cite{HH92}.
The entropy and  temperature of these black holes depend on
$\beta$ (see \cite{HW92}).
For the extremal black holes ( $r_{+} = r_{-} $ ) the entropy is finite for 
$\beta = 0$ and is zero for $\beta \neq 0$, and the temperature is zero for 
$\beta < 1$, is
finite ( the same as for a Schwarzschild black hole)  for $\beta = 1$,  and
is infinite for $\beta > 1$.
Holzhey and Wilczek\cite{HW92} showed that, for $\beta > 1$, infinite potential
barriers form around extremal charged dilaton black holes leading to the 
interpretation that these extremal black holes behave like 
elementary particles. They discussed that the extremal charged dilaton 
black holes  appear to be  extended spherical objects for $\beta < 1$ and 
 remarked that one can hardly refrain from observing that a string is the 
intermediate case between a spherical membrane and a point.
Horne and Horowitz\cite{HH92} and Shiraishi\cite{Shi92} obtained charged rotating dilaton
black hole solutions for small values of the rotation parameter. They
calculated the gyromagnetic ratio for these black holes and  found that it
depends on the free parameter $\beta$.

The  energy associated with the Schwarzschild and the RN
 black holes is well discussed in the  literature (see \cite{Mol58}-\cite{RN}
) and references  therein). The energy of the Schwarzschild black hole is 
confined to its interior  whereas the energy of the RN black hole is
shared by its interior as well as  exterior.
One of the present authors and Parikh\cite{VP93} investigated the energy of
a static spherically symmetric charged dilaton black hole for $\beta = 1$
and found the interesting result that, similar to the case of the Schwarzschild
black hole and unlike the RN black hole, the entire energy
is confined to its interior with no energy shared by the exterior
of the black hole. As  several
features of the charged dilaton black holes depend crucially on the coupling
parameter $\beta$, it is of  interest to study the energy
associated with  charged dilaton black holes for arbitrary values of 
$\beta$ to see $(a)$ what  the energy distribution 
 is for $\beta < 1$ as well as for $\beta >1$
, and $(b)$ whether or not the energy is confined to  the black hole interior
 for any  other value of $\beta$ ( apart from $\beta = 1$ ).

Since the outset of the general theory of relativity several prescriptions
for calculating the energy and momentum have been proposed, e.g. the energy-
momentum complex of Einstein\cite{Mol58}
\be
\Theta_i^{\ k} =  \frac{1}{16 \pi} H^{\ kl}_{i,\ \ l} ,
\ee
where
\be
H_i^{\ kl}  = \frac{g_{in}}{\sqrt{-g}}
         \lt[-g \lt( g^{kn} g^{lm} - g^{ln} g^{km}\rt)\rt]_{,m} \ .
\ee
$H_i^{\ kl}$ is antisymmetric in its contravariant indices. Latin indices take
values from $0$ to $3$ ($x^0$ is the time coordinate). $\Theta_i^{\ k}$, given 
by $ (10) $, satisfies the local conservation laws
\be
\frac{\partial \Theta_i^{\ k}}{\partial x^k} = 0 ,
\ee
where
\be
\Theta_i^{\ k} = \sqrt{-g} \lt(T_i^{\ k} + \vartheta_i^{\ k}\rt).
\ee
$T_i^{\ k}$ is the symmetric energy-momentum tensor due to matter and all
nongravitational fields and $\vartheta_i^{\ k}$ is the energy-momentum 
pseudotensor due to the gravitational field only.
The energy and momentum components are given by
\be
P_i = \frac{1}{16 \pi}
     \int \int \int {H_i^{\ 0\alpha}}_{,\alpha} \ dx^1\ dx^2\ dx^3 ,
\ee
where the Greek index $\alpha$ takes values from $1$ to $3$. $P_0$ stands for the 
energy (say $E$), and $P_1, P_2, P_3 $ are the momentum components.

The Einstein prescription for obtaining the energy and momentum associated with
asymptotically flat spacetimes gives meaningful result if calculations are
carried out in those coordinates in which the metric $g_{ik}$ approaches
the Minkowski metric $\eta_{ik}$ at great distance from the
system under investigation. These coordinates are usually called 
quasi-Cartesian or quasi-Minkowskian. Transforming the line 
element $(3)$ to these coordinates according to
\bea
x &=& r\ sin\th \  cos\varphi , \nn\\
y &=& r\ sin\th \  sin\varphi ,  \nn\\
z &=& r\ cos\th ,
\eea
one gets
\be
ds^2\ = B dt^2 - D (dx^2+dy^2+dz^2) - \frac{A-D}{r^2} (x
dx+y dy +z dz)^2.
\ee
The determinant $g  \equiv\  \vline g_{ik} \vline $ is 
\be
g = - \lt(  \frac{r}{r-r_{-}}  \rt)^{2(\sigma-1)} ,
\ee
and the nonvanishing contravariant components of the metric tensor are
\bea
g^{00} &=& \frac{r^{\sigma+1}}
              {(r-r_{-})^{\sigma} (r - r_{+})},     \nn\\
g^{11} &=& \frac{(r-r_{-})^{\sigma-1}}{r^{\sigma+3}}
      \lt[ x^2 \lt( rr_{-} + rr_{+} - r_{-} r_{+} \rt) - r^4   \rt],  \nn\\
g^{22} &=& \frac{(r-r_{-})^{\sigma-1}}{r^{\sigma+3}}
      \lt[ y^2 \lt( rr_{-} + rr_{+} - r_{-} r_{+} \rt) - r^4   \rt],  \nn\\
g^{33} &=& \frac{(r-r_{-})^{\sigma-1}}{r^{\sigma+3}}
      \lt[ z^2 \lt( rr_{-} + rr_{+} - r_{-} r_{+} \rt) - r^4   \rt],  \nn\\
g^{12} &=&  \ \frac{x y (r-r_{-})^{\sigma-1}} {r^{\sigma+3}}
            \lt[ r \lt( r_{-} + r_{+} \rt) - r_{-} r_{+}  \rt],  \nn\\
g^{23} &=&  \ \frac{y z (r-r_{-})^{\sigma-1}} {r^{\sigma+3}}
            \lt[ r \lt( r_{-} + r_{+} \rt) - r_{-} r_{+}  \rt] , \nn\\
g^{31} &=&  \ \frac{z x (r-r_{-})^{\sigma-1}} {r^{\sigma+3}}
            \lt[ r \lt( r_{-} + r_{+} \rt) - r_{-} r_{+}  \rt]. 
\eea
We are interested in  calculating the energy and therefore the required components
of $ H_i^{\ kl}$ are
\bea
H_0^{\ 01} &=& \frac{2x}{r^4} 
           \lt[r\lt(\sigma r_{-} + r_{+}\rt) - \sigma r_{-} r_{+} \rt], \nn\\
H_0^{\ 02} &=& \frac{2y}{r^4} 
           \lt[r\lt(\sigma r_{-} + r_{+}\rt) - \sigma r_{-} r_{+} \rt], \nn\\
H_0^{\ 03} &=& \frac{2z}{r^4} 
           \lt[r\lt(\sigma r_{-} + r_{+}\rt) - \sigma r_{-} r_{+} \rt]. 
\eea
By using $(19)$ with $(9)$ in $(14)$, applying the Gauss theorem, and then evaluating the 
integral over the surface of a sphere of radius $r$, we get
\be
E(r)\ =\ M\ - \ \frac{Q^2}{2r} \lt( 1 - \beta^2 \rt).
\ee
Thus we find that, like several other features of the charged
dilaton black holes, the energy distribution depends on the value of the
coupling parameter $\beta$. $\beta = 0$ in $(20)$ gives the energy distribution
in the RN field\cite{RN}. 
In the present investigation we find that {\em {only for}} $\beta = 1$
is the  energy  confined to its interior, and that for all other values of 
$\beta$ the energy is shared by the  interior and exterior of the black holes.
The total energy of the charged dilaton black holes is independent of 
 $\beta$ and is given by the mass parameter of the black hole.
With increasing radial distance, $E(r)$
increases for $\beta = 0$ (RN metric) as well as for $\beta<1$, decreases for
$\beta>1$, and remains constant for $\beta = 1$.

This work was presented in the Spanish Relativity Meeting in 1994
 and it has appeared in the Proceedings in brief \cite{CV95}.

\begin{flushleft}
{\bf Acknowledgments}
\end{flushleft}
One of us (KSV) thanks the Basque Government and the Tata
Institute of Fundamental Research for financial support.
This work was  supported in part by the Universidad del Pa\'{\i}s
Vasco under contract UPV 172.310 - EA062/93.
We thank A. Ach\'{u}carro, J. M. Aguirregabiria, and I. Egusquiza 
for  discussions.


\newpage

\begin{thebibliography}{99}
\setlength{\parskip}{0.32ex}

\bibitem{GM88}
        G. W. Gibbons and K. Maeda, {\it Nucl. Phys.} {\bf B298}, 741 (1988).
\bibitem{GHS91}
        D. Garfinkle, G. T. Horowitz and A. Strominger,{\it Phys. Rev.} 
        {\bf D43}, 3140 (1991) ; Erratum : {\it Phys. Rev.}{\bf D45}, 3888 
        (1992). 
\bibitem{HH92} 
        J. H. Horne and G. T. Horowitz, {\it Phys. Rev.} {\bf D46}, 1340 
        (1992). 
\bibitem{Shi92}
        K. Shiraishi, {\it Phys. Lett.} {\bf A166}, 298 (1992) . 
\bibitem{GH93} 
        R. Gregory and J. A. Harvey, {\it Phys. Rev.} {\bf D47},
        2411 (1993). 
\bibitem{HS92} 
       J. A. Harvey and A. Strominger, Quantum aspects of black holes,
       preprint EFI-92-41,  hep-th/9209055. 
\bibitem{STW91} 
       A. Shapere, S. Trivedi and F. Wilczek, {\it Mod. Phys.Lett.} {\bf A6},
       2677 (1991). 
\bibitem{Pre91}
       J. Preskill, P. Schwarz, A. Shapere, S. Trivedi and F. Wilczek, {\it 
       Mod. Phys. Lett.} {\bf 6}, 2353  (1991). 
\bibitem{MS94} 
       T. Maki and K. Shiraishi, {\it Class. Quant. Grav.} {\bf 11}, 227 
	   (1994). 
\bibitem{HW92}
       C. F. E. Holzhey and F. Wilczek, {\it Nucl. Phys.} {\bf B380}, 447 
       (1992). 
\bibitem{Dow94}
       F. Dowker, J. P. Gauntlett, D. A. Kastor and J. Traschen,
       {\it Phys. Rev. } {\bf D49},  2909  (1994). 
\bibitem{Mol58}
       C. M\o ller, {\it Ann. Phys. (NY)} {\bf 4}, 347 (1958).  
\bibitem{RN}  
       K. P. Tod. {\it Proc. R. Soc. Lond.} {\bf A388}, 467 (1983); 
       K. S. Virbhadra, {\it Phys. Rev.} {\bf D41}, 1086 (1990);
                        {\it Phys. Rev.} {\bf D42}, 2919 (1990); 
       F. I. Cooperstock and S. A. Richardson, {\it in Proc. 4th Canadian Conf.
          on General  Relativity and Relativistic Astrophysics ( World 
          Scientific,  Singapore, 1991)}; 
       A. Chamorro and K. S. Virbhadra, Pramana-J.Phys. {\bf 45}, 181 1995, 
          hep-th/9406148; 
       J. M. Aguirregabiria, A. Chamorro, and K. S. Virbhadra, {\it Energy and 
          angular momentum   of charged rotating black holes}, preprint
          gr-qc/9501002, Gen. Relativ. Gravit. (to appear). 
\bibitem{VP93}
        K. S. Virbhadra and J. C. Parikh, {\it Phys. Lett.} {\bf  B317}, 312 
          (1993) . 
\bibitem{CV95}
        A. Chamorro and K. S. Virbhadra, in {\it Inhomogeneous Cosmological 
        Models, Eds. A Molina and J M M Senovilla} (World Scientific, Singapore 1995)
        p230, gr-qc/9602005.

\end{thebibliography}
\end{document}